\documentclass[twoside]{article}
\usepackage{amsfonts,amssymb,amsmath,latexsym,cite}
\usepackage{epsfig,graphicx}
\usepackage{indentfirst}
\usepackage{cite}
\newcommand{\mytitle}[1]{\large \sc #1 \\}
\newcommand{\avtor}[1]  {\large \it #1 \\}

mm  \leftmargin 5pt
\newtheorem{demo}{Statement}

\makeatletter \@addtoreset{equation}{section}

\@addtoreset{demo}{section}  \makeatother

\begin{document}

\begin{center}

\mytitle{Berry phases for  3D Hartree type equations with a
quadratic potential and  a uniform magnetic field}
\bigskip

\avtor {F.~N. Litvinets $^{*}$\footnote{e-mail:
litvinets@mph.phtd.tpu.edu.ru}, A.~V.
Shapovalov$^{*\#}$\footnote{e-mail: shpv@phys.tsu.ru}, and A.~Yu.
Trifonov $^*$\footnote{e-mail: trifonov@mph.phtd.tpu.edu.ru} }

$^{*}$ {\it Laboratory of Mathematical Physics\\
Mathematical Physics Department,\\
Tomsk Polytechnic University,\\
30, Lenin ave., Tomsk, 634050, Russia }\vskip 0.4cm
$^{\#}$ {\it Theoretical Physics Department,\\
Tomsk State University, \\
36, Lenin ave., Tomsk, 634050, Russia}
\end{center}

\begin{abstract}

A countable set of asymptotic space -- localized solutions is
constructed by the complex germ method in the adiabatic
approximation for 3D Hartree type equations with a quadratic
potential. The asymptotic parameter is $1/T$, where $T\gg1$ is the
adiabatic evolution time.

A generalization of the Berry phase of the linear Schr\"odinger
equation is formulated for the Hartree type equation. For the
solutions constructed, the Berry phases are found in explicit form.

\end{abstract}

\section{Introduction}

The method of semiclassical asymptotics   developed in \cite
{Litv:LTS1, Litv:LTS3, shapovalov:BTS1, shapovalov:BTS2} is used to
construct and study the solutions of a many-dimensional
non-stationary Hartree type equation with a nonlocal nonlinearity
and the coefficients adiabatically  varying in time. By adiabatic
evolution of parameters we mean their relative small  changes in
some characteristic time (adiabatic evolution time). In particular,
for parameters periodic in time, the characteristic time is the
period.

In the time of adiabatic evolution, a quantum system  returns to its
initial state and the wave function gains only a phase factor. M.
Berry has revealed \cite{shapovalov:BERRY} that the total phase
contains, along with the dynamic part  known from the  Born--Fock
adiabatic theorem \cite{shapovalov:VINI, Born, Messia}, a geometric
additive component (geometric phase (GP)). This summand in the total
phase is known as the  adiabatic phase or the Berry phase. The
adiabatic phase is closely connected with the Floquet problem for
systems of differential equations with periodic coefficients
\cite{Floke}. In  quantum mechanics, geometric phases  are well
investigated for the linear Schr\"odinger equation
\cite{shapovalov:VINI, Moore, shapovalov:KLYSHKO, Biswas}. In
classical mechanics, a Hannay angle is introduced for nearly
integrable Hamiltonian systems with adiabatically varying
parameters. The Hannay angle is related to the Berry phase
\cite{Berry85} if the Hamiltonian system corresponds  to the
considered quantum system. The Hannay angle is defined as an
additional term to the dynamic contribution in the ``angle''
variable when the Hamiltonian system is described in terms of
``action-angle'' variables \cite{Hannay1,Arnold}.

Geometric phases are observable \cite{Tomita, Suter, Antaramian,
Chiao, Scala, Bitter} and they show up in various physical phenomena
\cite{Baily, Delacretaz, Stern}. It is considered that in quantum
calculations \cite{Deutsch,Preskill}, intensely developing at
present, the  Berry phase  might be used in some types of quantum
gate, the so-called geometric gates (GGs)\cite {Ekert}. The last
ones offer some advantages over the conventional (non-geometric)
phase gates as the GGs have the greater fault tolerance. An example
of an engineering  embodiment of geometric gates based on the
nuclear magnetic resonance is given in \cite{Jhonatan}. To build
hardware for a quantum computer systems of cooled ions can be used
\cite {Preskill}. In such a system, each ion caries a qubit, and the
execution of a logic operation (gate) is governed by an external
electromagnetic field created by laser radiation, a magnetic field
source, etc. It should be noted that the Paul traps can be described
by the potential of a harmonic oscillator \cite{Scala, Preskill}.
One should also take into account the collective coupling of ions as
the ion states depend not only on the external field, but also on
the collective behavior of the ions \cite{Preskill}. A consideration
of the coupling between parts of a system naturally leads to
nonlinear models. An example of such a system might be the
Bose--Einstein condensate (BEC)\cite{shapovalov:CORNELL,Pitaevskii}
models of which are used the Gross-Pitavskii equation
\cite{shapovalov:GROSS,shapovalov:PITAEVSKII-1}. This is an argument
in favor of to study such models taking in mind their applications.

The geometric phases in nonlinear systems were studied in
\cite{Fazakon1,Fazakon3} where the Berry phase was found for states
describing a two-component BEC in an external field. In such
systems, the Berry phase arises  when the parameters responsible for
the coupling of the condensate subsystems change slowly.

Here we continue our investigations started in
\cite{KievPhase,LfTS1} where geometric phases were studied for a one
dimensional Hartree type equation. The aim of this work is to find
the Berry phase for a many-dimensional Hartree type equation of the
following form:

\begin{eqnarray}
&&\bigg\{ -i\hbar\partial_t
+\widehat{\mathcal H}_{\varkappa}(R(t),\Psi(t))\bigg\}\Psi =0,\label{pgau1}\\
&&\widehat{\mathcal H}_{\varkappa}(R(t),\Psi(t))=\widehat{\mathcal
H}(R(t))+\varkappa\widehat V(R(t),\Psi(t)),\label{phbb07}\\
&&\displaystyle\widehat{\mathcal H}(R(t))=
\frac1{2m(t)}\bigg(\hat{\vec{p}}-\frac{e}{c}
\vec{A}(\vec{x},t)\bigg)^2+\frac{\rho(t)\big(\langle\vec x,
\hat{\vec p}\rangle+\langle\hat{\vec p}, \vec x\rangle\big)}2
+\frac{k(t) \vec{x}^2}{2},\cr &&\displaystyle\widehat
V(R(t),\Psi(t))=\frac12
  \int\limits_{\mathbb{R}^3}d\vec y\,
{\Big[a(t)\vec{x}^2+2b(t)\langle\vec{x},\vec{y}\rangle+c(t)\vec{y}^2\Big]
|\Psi(\vec{y},t)|^2},\label{V}\\ &&\displaystyle
\vec{A}(\vec{x},t)=\frac12[\vec{H}(t),\vec{x}].\nonumber
\end{eqnarray}
Here, $a(t)$, $b(t)$, and $c(t)$ are parameters of the nonlocal
potential; $\varkappa$ is a nonlinearity parameter; $m(t)$, $k(t)$,
and $\rho(t)$ are time-dependent parameters of the system,
$\vec{H}(t)$ is an external magnetic force; $\langle\vec a,\vec
b\rangle$ and $[\vec a,\vec b]$ are, respectively, the scalar and
the vector product in $\mathbb{R}^3$. The nonlinear Hamiltonian
$\widehat{\mathcal H}_{\varkappa}$ depends on time via the set of
parameters $R(t)$=($m(t)$, $k(t)$, $\rho(t)$, $\vec{H}(t)$, $a(t)$,
$b(t)$, $c(t)$).

The states with nontrivial geometric phases exist for quantum
systems with Hamiltonians quadratic in variables and derivatives.
Such Hamiltonians, in particular, include the external field of a
harmonic oscillator and a uniform magnetic field \cite {ty94}.

The  magnetic field reduces the symmetry of a quantum system,
removing its degeneration, which in turn can result, for instance,
in a normal Zeeman effect \cite {Landau3}, etc.

The study of geometric phases for nonlinear equations is a
complicated mathematical problem since to construct the phase
requires methods for solving many-dimensional nonlinear equations
with variable coefficients in an appropriate class of functions. The
well-known Inverse Scattering Transform method
\cite{ZAKHAROV,Calogero,Newell} is applied mainly to  $(1+1)D$ and
$(1+2)D$ nonlinear equations with constant coefficients, and only
soliton solutions can be constructed in explicit form. The symmetry
analysis \cite{Ovsjannikov,Pukhnachov,Anderson,Olver,Nikitin,Gaeta}
allows one to study systems possessing high symmetry, but the
evaluation of a symmetry is inconvenient if the equation contains
nonlocal terms. The nonlinear problems of the above class could be
effectively solved by the method developed in \cite{Litv:LTS1,
Litv:LTS3, shapovalov:BTS1, shapovalov:BTS2} where asymptotic
solutions (in some cases, exact solutions) are constructed for a
nonlinear Hartree type equation which is the Gross-Pitaevskii
equation with a nonlocal nonlinearity.

This paper is organized as follows. In the first section, necessary
designations and definitions are introduced for the Berry phase. In
Section 2, a method of seeking an exact solution for a Hartree type
equation with a quadratic Hamiltonian is briefly described. Section
3 presents a solution of the spectral problem for an instantaneous
Hartree type operator. The solution is used to extract the dynamic
phase from the overall phase. In Section 4, solutions of the Hartree
type equation are constructed in the adiabatic approximation and the
corresponding Berry phases are found. In Conclusion, the results and
related problems are discussed.

\section{The Berry phase for a nonlinear equation }
In this work, we find the Berry phase using the approach developed
in \cite{DodKlMan,ty94} for linear quantum equations. This approach
is based on seeking an exact (or approximate) solution of the Cauchy
problem for equation (\ref{pgau1}):
\begin{equation}
\Psi\big|_{t=0}=\psi_{\nu}(\vec x,R(0)).\label{Ber2_3}
\end{equation}
which then is expanded in an adiabatic parameter. Here the functions
$\psi_{\nu}(\vec x,R(0))$ are determined by the spectral problem for
the instantaneous Hartree type operator
\begin{equation}
\widehat{\mathcal H}_\varkappa(R(t))\psi_{\nu}(\vec x,R(t))=
E_\nu(R(t))\psi_{\nu}(\vec x,R(t)).\label{Ber2_2}
\end{equation}

Assume that the parameters $R(t)$ are T-periodic and slowly vary
with time\footnote{A system is assumed to be adiabatic if the
following condition is fulfilled:
\begin{equation}
\max_{i=\overline{1,n}}\dot{R_{i}}\frac{T}{R_{i}}\ll 1,\label{phb07}
\end{equation}
where $R_{i}$ are parameters of the Hamiltonian (see, e.g.,
\cite{Landau}).}. Following the adiabatic theorem in (linear)
quantum mechanics \cite{Messia}, we seek a solution of the Cauchy
problem (\ref{pgau1}), (\ref{Ber2_3}) in the form
\begin{equation}
\Psi(\vec x,t)=\exp[{i}\phi_{\nu}(t)]\psi_{\nu}(\vec
x,R(t))+O\bigg(\displaystyle\frac1T\bigg). \label{Ber2_4}
\end{equation}
Then for $t=T$, taking into account that $R(T)=R(0)$, we have
\begin{equation} \Psi(\vec x,
T)=\exp[{i}\phi_{\nu}(T)]\Psi(\vec
x,0)+O\bigg(\displaystyle\frac1T\bigg). \label{Ber2_4a}
\end{equation}
Let us rewrite the phase $\phi_{\nu}(T)$ as
\begin{equation}
\phi_{\nu}(T)=\delta_{\nu}(T)+\gamma_{\nu}(T), \label{Ber2_5}
\end{equation}
where $\delta_{\nu}(T)$ is the dynamic phase, which is fined by the
relation
\begin{equation}
\delta_{\nu}(T)=-\frac{1}{\hbar}\int\limits_{0}^{T}E_\nu(R(t))dt.\label{phb01}
\end{equation}
Following \cite{shapovalov:BERRY} (see also \cite{Moore, Biswas,
LfTS1}), we call the phase $\gamma_{\nu}(T)$ the adiabatic Berry
phase or geometric phase for equation (\ref{pgau1}). The Berry phase
for the linear Schr\"odinger equation ($\varkappa=0$ in
(\ref{phbb07})) is determined by
\begin{equation}
\gamma_{\nu}(T)=i\int\limits_{0}^{T}\langle\psi_{\nu}(\vec x,R(t))|
\dot{\psi}_{\nu}(\vec x,R(t))\rangle dt.\label{phb04}
\end{equation}
Formula (\ref {phb04}) is equivalent to (\ref {Ber2_5}) in the
linear case and requires an additional substantiation for nonlinear
equations.

\section{Method of semiclassically concentrated functions }

Consider exact solutions for equation (\ref{pgau1}) following
\cite{Litv:LTS1, Litv:LTS3, shapovalov:BTS1, shapovalov:BTS2}.  From
these solutions we find solutions of the form (\ref{Ber2_4}) and
obtain the Berry phase.

Define the mean value for a linear operator $ \widehat A $  in a
state $\Psi (t)$ as
\begin{equation}
\label{gau1a} \langle\widehat
A(t)\rangle=A_\Psi(t,\hbar)=\frac{1}{\|\Psi(t)\|^2}
\langle\Psi(t)|\widehat A| \Psi(t)
\rangle=\frac{1}{\|\Psi(t)\|^2}\int\limits_{{\mathbb R}^3} d \vec x
{\Psi}^* (\vec x,t,\hbar)\widehat A(t)\Psi (\vec x,t,\hbar),
\end{equation}
where $\|\Psi (t)\|^2={\langle \Psi (t) |\Psi (t) \rangle}$. For the
solutions $ \Psi (t) $ of equation (\ref {pgau1}) we have
\begin{eqnarray}
\label{gau1b} &\displaystyle\frac{d\langle\widehat A(t)\rangle}{d
t}=\Big\langle\frac{\partial \hat A(t)}{\partial t}
\Big\rangle+\frac{i}{\hbar}\langle [\widehat{\mathcal
H}_\varkappa(t,\Psi(t)),\widehat A(t)]_{-} \rangle,
\end{eqnarray}
where $[\hat A,\hat B]_{-}=\hat A\hat B-\hat B\hat A$ is the
commutator of  linear operators  $\hat A$ and $\hat B$.

As in the linear case, we call  equation (\ref {gau1b}) an Ehrenfest
equation. From (\ref{gau1b}) with $\widehat A=1 $ it follows, in
particular, that the norm of a solution of the equation
(\ref{pgau1}) does not depend on time, i.e.
\[
\| \Psi (\vec x, t) \| ^2 = \|\Psi (\vec x, 0) \| ^2 = \|\Psi \|^2.
\]
Then it is convenient to go from the constant $ \varkappa $ to a
constant $\tilde\varkappa =\varkappa \|\Psi \| ^ 2 $.

Let us write down the Ehrenfest equation for mean values of the
operator  $\hat {\mathfrak g}$:
\begin{eqnarray}
\hat{\mathfrak g}=\bigg({\hat z}_{j},\displaystyle\frac12(
 \Delta\hat {z}_{k}\Delta\hat {z}_{l}+\Delta\hat {z}_{l}\Delta\hat {z}_{k});
\quad j,k,l=\overline{1,6}\bigg)^\intercal;\quad
 \hat z_m=\hat{p}_m,\,\quad \hat z_{m+3}=x_m,\,m=\overline{1,3};\label{kvs7q}
\end{eqnarray}
where  $\Delta \hat z_k=\hat z_k-z_{k\Psi}(t,\hbar)$, $z_{k\Psi}(t,
\hbar)=\langle\hat z_k\rangle$. As a consequence, we obtain for the
first- and second-order moments
\begin{equation}
\left\{
\begin{array}{l}\dot z_\Psi = J{\mathfrak H}_z({\mathfrak g}_\Psi,R(t)),\\[8pt]
 \dot\Delta_{2\Psi}=J{\mathfrak H}_{zz}(R(t))\Delta_{2\Psi}-\Delta_{2\Psi}{\mathfrak H}_{zz}(R(t))J,
 \quad\Delta_{2\Psi}^\intercal=\Delta_{2\Psi},
\end{array}\right.\label{bbst2.6a1}
\end{equation}

\begin{eqnarray}
&&{\mathfrak H}_z( {\mathfrak g}_{\Psi},R(t))=\left(
 \begin{array}{c}\displaystyle\frac{\vec p_\Psi}{m(t)}+\rho(t) \vec x_\Psi
 -\displaystyle\frac{e}{2m(t)c}[\vec H(t),\vec x_\Psi]\\[4pt]
 k_1(t)\vec x_\Psi+\rho(t) \vec p_\Psi+\displaystyle\frac{e}{2m(t)c}[\vec H(t),
 \vec p_\Psi]+
 \displaystyle\frac{e^2}{4m(t)c^2}[\vec H(t),[\vec x_\Psi,\vec
 H(t)]]\end{array}\right),\nonumber\\[8 pt]
&&\qquad\qquad\qquad\qquad\qquad {\mathfrak
H}_{zz}(R(t))=\left(\begin{array}{cc}
   \mathfrak H_{pp}(R(t)) &\mathfrak H_{px}(R(t)) \\[4pt]
   \mathfrak H_{px}^\intercal(R(t)) &\mathfrak H_{xx}(R(t))\end{array}\right),
   \nonumber\\
&& \mathfrak
H_{pp}(R(t))=\displaystyle\frac{1}{2m(t)}\mathbb{E},\quad
\mathfrak H_{px}(R(t))=\displaystyle\frac{e}{2m(t)c}\left(%
\begin{array}{ccc}
  \displaystyle\frac{2m(t)\rho(t)c}{e} & H_3(t) & -H_2(t) \\[4pt]
  -H_3(t) & \displaystyle\frac{2m(t)\rho(t)c}{e} & H_1(t) \\[4pt]
  H_2(t) & -H_1(t) & \displaystyle\frac{2m(t)\rho(t)c}{e}
\end{array}\right),\label{Hzz}\\[8pt]
&&{\mathfrak H}_{xx}(R(t))=\displaystyle\frac{e^2}{4m(t)c^2}\times\cr&&\times\left(%
\begin{array}{ccc}
  \vec{H}^2(t)-H_1^2(t)+\displaystyle\frac{4m(t)\tilde{k}(t)e^2}{c^2} & -H_1(t)H_2(t) & -H_1(t)H_3(t) \\[4pt]
  -H_2(t)H_1(t) & \vec{H}^2(t)-H_2^2(t)+\displaystyle\frac{4m(t)\tilde{k}(t)e^2}{c^2} & -H_2(t)H_3(t) \\[4pt]
  -H_3(t)H_1(t) & -H_2(t)H_1(t) & \vec{H}^2(t)-H_3^2(t)+\displaystyle\frac{4m(t)\tilde{k}(t)e^2}{c^2}
\end{array}\right).\nonumber
\end{eqnarray}
Here $\mathbb{E}$ is an identity matrix, $B^\intercal$ is the
transpose to the matrix $B$;
$k_1(t)=k(t)+\tilde\varkappa(a(t)+b(t))$,\linebreak $\tilde
k(t)=k(t)+\tilde\varkappa a(t)$.

The matrix of the second centered moment $\Delta_{2\Psi}$ has the
form
\begin{equation}
\label{shapovalov:1.12} \Delta_{2\Psi}(t,\hbar)=\left (
\begin{array}{cc}
\sigma_{pp}^{\Psi}(t,\hbar)& \sigma_{px}^{\Psi}(t,\hbar)\\[4pt]
\sigma_{xp}^{\Psi}(t,\hbar) &\sigma_{xx}^{\Psi}(t,\hbar)
\end{array} \right ),
\end{equation}
where
\begin{eqnarray*}
&&\sigma_{pp}^{\Psi}(t,\hbar)=\|\sigma_{p_k
p_l}^{\Psi}(t,\hbar)\|_{3\times 3}=\|\langle
\Delta\hat p_k \Delta\hat p_l\rangle)\|_{3\times 3}, \\[4 pt]
&&\sigma_{xx}^{\Psi}(t,\hbar)=\|\sigma_{x_k
x_l}^{\Psi}(t,\hbar)\|_{3\times 3}=\|\langle \Delta x_k
\Delta x_l\rangle\|_{3\times 3},\qquad\qquad\qquad \qquad\qquad    k,l=\overline{1,3}, \\[2 pt]
&&\sigma_{xp}^{\Psi}(t,\hbar)=\|\sigma_{x_k
p_l}^{\Psi}(t,\hbar)\|_{3\times 3}=
\Big\|\displaystyle\frac{1}{2}\langle \Delta x_{k} \Delta \hat
p_l+\Delta\hat p_l \Delta x_k \rangle\Big\|_{3\times 3}.
\end{eqnarray*}

We call the system of equations (\ref{bbst2.6a1}) the second order
{\em Hamilton-Ehrenfest system (HES)} corresponding to equation
(\ref{pgau1}).

The matrix of the second centered moments\footnote {The subscript
$\Psi$ can be omitted} $\Delta_2$ can be rewritten
\begin{equation}
\Delta_2(t)= A(t)\Delta_2(0) A^+(t),\label{secmom}
\end{equation}
where $A(t)$ is the fundamental matrix of the system in variations
\begin{equation}
 \dot A= J {\mathfrak H}_{zz}(R(t)) A,
\qquad A(0)=\Bbb I.\label{matvar}
\end{equation}
We denote by ${\mathfrak g}(t,{\mathfrak C})=(Z(t,{\mathfrak C}),
\Delta_2(t,{\mathfrak C}))$ the general solution of the system
(\ref{bbst2.6a1}), where  ${\mathfrak C}$ are integration constants.

Let us seek a solution of equation (\ref{pgau1}) in terms of the
anzats
\begin{equation}
\Psi(\vec x,t,\hbar)=\varphi\Bigl(\frac{\Delta\vec
x}{\sqrt{\hbar}},t,
  \sqrt{\hbar}\Bigr)\exp\Bigl[{\frac{i}{\hbar}
  \Bigl(S(t,{\hbar})+\langle\vec P(t,{\hbar}), \Delta\vec x\rangle\Bigr)}\Bigr].
  \label{gau4}
\end{equation}
Here the function $\varphi(\vec\xi ,t,\sqrt{\hbar} )$ belongs to the
Schwartz space ${\Bbb S}$ in the variable $\vec\xi=\Delta\vec
x/\sqrt\hbar$   and regularly depends on $\sqrt{\hbar}$; $\Delta
\vec x=\vec x-\vec X(t,{\hbar})$. The real function $S(t,\hbar)$ and
vector function  $Z(t,{\hbar})=(\vec P(t,\hbar),\vec X(t,{\hbar}))$
are to be determined.

Let us expand the operators entering equation  (\ref{pgau1}) as
Taylor series in  $\Delta\vec x=\vec x-\vec x_\Psi(t,\hbar)$,
$\Delta \vec y=\vec y-\vec x_\Psi(t,\hbar)$, and  $\Delta \hat{\vec
p}=\hat{\vec p}-\vec p_\Psi(t,\hbar)$. Then equation (\ref{pgau1})
takes the form
\begin{eqnarray}
&&\displaystyle\{ -i\hbar\partial_t +{\mathfrak H}(\Psi,t)+
\langle{\mathfrak H}_z(\Psi,t),\Delta\hat z\rangle
+\frac12\langle\Delta\hat z,{\mathfrak H}_{zz}(t)\Delta\hat z\rangle\}\Psi=0, \label{gau6}\\
&&\displaystyle{\mathfrak H}(\Psi,t)={\mathfrak H}({\mathfrak
g}_{\Psi}(t,\hbar),R(t))=\frac{\vec
p_\Psi^2(t,\hbar)}{2m(t)}+\rho(t)\langle\vec x_\Psi, \vec
p_\Psi\rangle+\frac{k_0(t)\vec
x_\Psi^2(t,\hbar)}{2}-\frac{e}{2m(t)c}\langle[\vec H(t),\vec
x_\Psi(t,\hbar)], \vec p_\Psi(t,\hbar)\rangle+\cr
&&\qquad+\frac{e^2}{8m(t)c^2} \bigg[\vec H^2(t)\vec
x^2_\Psi(t,\hbar)-(\langle \vec H(t),\vec
x_\Psi(t,\hbar)\rangle)^2\bigg]
+\displaystyle\sum_{k=1}^3\frac{\tilde\varkappa}2
c(t)\,\sigma_{x_k x_k}^{\Psi}(t,\hbar), \nonumber\\
&&{\mathfrak H}_z(\Psi,t)={\mathfrak H}_z({\mathfrak
g}_{\Psi}(t,\hbar),R(t)),\qquad{\mathfrak H}_{zz}(t)={\mathfrak
H}_{zz}(R(t)).\nonumber
\end{eqnarray}
Here $k_0(t)=k(t)+\tilde\varkappa(a(t)+2b(t)+c(t))$; the vector
${\mathfrak H}_z({\mathfrak g}_{\Psi},R(t))$ and matrix ${\mathfrak
H}_{zz}(R(t))$ are given in (\ref{Hzz}).

Let us relate the nonlinear  equation   (\ref{gau6}) with the linear
equation that is obtained from (\ref{gau6}) by formal replacement of
the mean values of the operators of coordinates, momenta, and
centered moments of the second order ${\mathfrak g}_\Psi(t,\hbar)$
by the general solution ${\mathfrak g}(t,{\mathfrak C})$  of the
Hamilton-Ehrenfest system (\ref{bbst2.6a1}):
\begin{eqnarray}
&&\displaystyle\{ -i\hbar\partial_t +{\mathfrak H}(t,\mathfrak C)+
\langle{\mathfrak H}_z(t,\mathfrak C),\Delta\hat z\rangle
+\frac12\langle\Delta\hat z,{\mathfrak H}_{zz}(t)\Delta\hat z\rangle\}\Phi=0, \label{gau6a}\\
&&{\mathfrak H}(t,\mathfrak C)={\mathfrak
H}(\Psi,t)\Big|_{{\mathfrak g}_{\Psi}(t)\to{\mathfrak
g}(t,{\mathfrak C})},\quad{\mathfrak H}_z(t,\mathfrak C)={\mathfrak
H}_z(\Psi,t)\Big|_{{\mathfrak g}_{\Psi}(t)\to{\mathfrak
g}(t,{\mathfrak C})}.\nonumber
\end{eqnarray}

We call equation  (\ref{gau6a}) {\em the associated linear
Schr\"odinger equation}. We can immediately verify that the function
\begin{equation}
\Phi_0(x,t,{\mathfrak C})=\mid0,t,{\mathfrak
C}\rangle=N_\hbar\Biggl(\frac{\det C(0)}{\det
C(t)}\Biggr)^{1/2}\exp{ \biggl\{\frac{i}{\hbar}\Big(S(t,{\mathfrak
C})+ \langle\vec P(t,{\mathfrak C}),\Delta\vec x\rangle+\frac
12\langle\Delta\vec x, Q(t)\Delta\vec x\rangle\Big)\biggr\}},
\label{spgau13a}
\end{equation}
where
\begin{equation}
S(t,{\mathfrak C})=\int\limits_0^t\big(\langle\vec P(t,{\mathfrak
C}),\dot {\vec{X}}(t,{\mathfrak C})\rangle-{\mathfrak
H}(t,{\mathfrak C})\big)dt. \label{gau13c}
\end{equation}
is a solution of equation (\ref{gau6a}). Here $Q(t)=B(t)C^{-1}(t)$,
and by $B(t)$ and $C(t)$ we have designated  the ``momentum'' and
``coordinate'' parts of the matrix solution of the system in
variations  (see \cite{Bagre}) corresponding to the linear equation
(\ref{gau6a}):
\begin{equation}
\left\{\begin{array}{l} \dot B=-{\mathcal H}_{xp}(t)B-{\mathcal H}_{xx}(t)C,\\
\dot C={\mathcal H}_{pp}(t)B+{\mathcal
H}_{px}(t)C.\vspace{-3pt}\end{array} \right.\label{kak19}
\end{equation}
Let us write down a matrix
\begin{equation}
\mathcal{A}(t)=\left(\begin{array}{l} B(t)\\[4pt]
C(t)\end{array}\right).\label{gau13b}
\end{equation}

The matrix $\mathcal{A}$ can be represented as
\begin{equation}
\mathcal{A}=(a_1, a_2, a_3),
\end{equation}
where  $\{a_k\}$ is a set of linearly independent vectors ---
solutions of the equation
\begin{equation}
\dot a_k=J{\mathfrak H}_{zz}(t)a_k, \label{gau13q.1}\quad
k=\overline{1,3}.
\end{equation}
An operator\footnote {we omit the dependence on $\hbar$ and
${\mathfrak C}$}
\begin{equation}
\hat a(t)= N_a\langle a(t),J^\intercal\Delta\hat z\rangle
\label{phaz1lst4.13}
\end{equation}
is a symmetry operator for equation  (\ref{gau6a}) if the vector
$a(t)$ is a solution of the system in variations
(\ref{gau13q.1})\cite{Bagre}.  Let $\hat a(t)$ and $\hat b(t)$ be
symmetry operators corresponding to the two solutions of the system
in variations,  $a(t)$ and  $b(t)$, respectively. Then it is easy to
verify that
\begin{equation}
[\hat a(t),\hat b(t)]_{-}=i\hbar N_aN_b\{a(t),b(t)\}=i\hbar
N_aN_b\{a(0),b(0)\}. \label{phaz1lst4.13'}
\end{equation}
The last equality in (\ref{phaz1lst4.13'}) is due to the Hamiltonian
form of system  (\ref{gau13q.1}). By braces we designate the
skew-scalar product of the two vectors $\{a,b\}=\langle
a,J^\intercal b\rangle$.

Assume that the system in variations  (\ref{gau13q.1}) admits a set
of  three, linearly independent complex solutions $a_k(t) = (\vec
W_k(t),\vec Z_k(t))^\intercal$ satisfying the skew-orthogonal
condition
\begin{equation}
\{a_k(t),a_l(t)\} = 0, \qquad \{a_k(t), a^*_l(t)\} = 2i\delta_{kl},
\qquad k,l = \overline{1,3}. \label{phaz1lst4.2a}
\end{equation}
Recall that  the $6$--vectors $a_k(t)$ and $a^*_k(t)$,
$k=\overline{1,3}$, serve as a symplectic basis in ${\mathbb
C}^{6}_a$, and the $3$-dimensional plane  $r^3(Z(t,{\mathfrak C}))$
with the basis $a_k(t)$ constitutes a complex germ on
$z=Z(t,{\mathfrak C})$ \cite{Mas2,BelDob}.

Setting $N_j=(\hbar)^{-1/2}$ in formula (\ref{phaz1lst4.13}), we
compare the vectors $a_j^*(t)$ with the ``creation'' operators $\hat
a_j^+(t)$, and the vectors $a_j(t)$ with the ``annihilation''
operators $\hat a_j(t)$. Then, taking into account
(\ref{phaz1lst4.13'}), we obtain for the operators  $\hat a_j^+(t)$,
$\hat a_j(t)$ the bosonic commutation relations
\begin{equation}
[\hat a_j(t),\hat a_k(t)]_{-}=[\hat a_j^+(t), \hat a_k^+(t)]_{-}=0,
\quad [\hat a_j(t),\hat a_k^+(t)]_{-}=\delta_{jk}, \quad
j,k=\overline{1,3}. \label{phaz1lst4.14}
\end{equation}

\begin{demo}
Let  $a_j(t)=(\vec W_j(t),\vec Z_j(t))$ be solutions of the problem
{\rm (\ref{gau13q.1}), (\ref{phaz1lst4.2a})}. Then the function
$\Phi_0(\vec x,t,{\mathfrak C})=|0,t,{\mathfrak C}\rangle$ {\rm
(\ref{spgau13a})} is a ``vacuum'' trajectory-coherent state
\begin{equation}
\hat a_j(t,{\mathfrak C})|0,t,{\mathfrak C}\rangle =0, \quad
j=\overline{1,3}, \label{phaz1lst4.15}
\end{equation}
for the associated linear Schr\"odinger equation {\rm(\ref{gau6a}).}
\end{demo}
{\bf Proof.} Applying the ``annihilation'' operator  $\hat a_j(t)$
to the function $|0,t\rangle$,  we find that
 $$
 \hat a_j|0,t\rangle
=|0,t\rangle [\langle\vec Z_j(t),Q(t)\Delta \vec x\rangle
-\langle\vec W_j(t),\Delta\vec x\rangle ].
$$
 It follows immediately that
(\ref{phaz1lst4.15}) is true as
\[Q(t)\vec Z_j(t)=B(t)C^{-1}(t)\vec Z_j(t)=\vec
W_j(t)\] according to the definition and properties of the matrix
 $Q(t)$.

Let us define now a countable set of states  $|\nu,t,{\mathfrak
C}\rangle$ (exact solution of equation  (\ref{gau6a})) as a result
of the action of the ``creation'' operators on the ``vacuum'' state
$|0,t,{\mathfrak C}\rangle$ (\ref{spgau13a}):
\begin{eqnarray}
&&\Phi_\nu({\vec x},t,{\mathfrak C})=|\nu,t,{\mathfrak
C}\rangle=\frac 1{\sqrt{\nu!}}(\hat a^+(t,{\mathfrak C}))^\nu
|0,t,{\mathfrak C}\rangle=\cr&&\quad=\prod_{k=1}^{3}\frac
1{\sqrt{\nu_k!}}(\hat a_k^+(t,{\mathfrak C}))^{\nu_k}
|0,t,{\mathfrak C}\rangle, \quad \nu=(\nu_1, \nu_2, \nu_3).
\label{phaz1lst4.16}
\end{eqnarray}
Notice that, in  view of  the explicit form of the ``vacuum'' state
(\ref{spgau13a})  and  symmetry operators (\ref{phaz1lst4.13}) for
the functions  $\Phi_\nu({\vec x},t)$ (\ref{phaz1lst4.16})
constituting the Fock basis, the following representation is true:

\begin{eqnarray}
&&\Phi_\nu({\vec x},t)=N_\nu\Phi_0({\vec x},t){\rm
He}_\nu\bigl(\vec{\zeta}(R,t),t\bigr).\label{phipol}
\end{eqnarray}
 Here,  ${\rm He}_\nu\bigl(\vec{\zeta}(R,t),t\bigr)$ are
 many-dimensional Hermite polynomials, which are set
by the matrix  $W(t)$ (see  \cite{Beitman2} for details),
\begin{eqnarray}
&&{\rm He}_\nu\bigl(\vec{\zeta}(R,t),t\bigr)={(-1)^{|\nu|}}
\biggl(\displaystyle\frac{\partial}{\partial\vec{\zeta}}-2W(t)\vec{\zeta}\biggr)^{\nu}\cdot1,\quad
N_\nu=\bigg(\displaystyle\frac{1}{\sqrt{2}}\bigg)^{|\nu|}\displaystyle\frac{1}{\sqrt{\nu!}};
\\
&&\vec{\zeta}(R,t)=\displaystyle\frac{-i}{\sqrt{\hbar}}(C^*(t))^{-1}\Delta\vec{x},\quad
W(t)=C^{+}(t)(C^{-1}(t))^\intercal.
\end{eqnarray}
The functions $\Phi_\nu(x,t,{\mathfrak C})$ are solutions of
 equation (\ref{gau6}) with a proper choice of
${\mathfrak C}$, such that the solutions ${\mathfrak g}(t,\mathfrak
C)$ of the Hamilton-Ehrenfest system  (\ref{bbst2.6a1})  coincide
with the corresponding mean values ${\mathfrak
g}_{\Phi_\nu}(t,\hbar,{\mathfrak C})$ for the states
(\ref{phaz1lst4.16}) (see \cite{LTS3}).

Let us designate this set of parameters by  $\overline{{\mathfrak
C}}_\nu$; then we have
\begin{equation}
\Psi_{\nu}(x,t,\hbar)=\Phi_\nu(x,t,\overline{{\mathfrak
C}}_\nu).\label{astohar}
\end{equation}
The subscript $\nu$ in  $\overline{{\mathfrak C}}_\nu$ shows that
for each function $\Psi_{\nu}(x,t,\hbar)$ there exists its own set
of the parameters  $\overline{{\mathfrak C}}_\nu$.

The construction of solutions (\ref{astohar}) uses solutions of two
auxiliary ordinary differential systems: the Hamilton--Ehrenfest
system (\ref{bbst2.6a1}) and  system in variations (\ref{gau13q.1}).

For an arbitrary time dependence of the coefficients $R(t)$, the
solutions of these systems are unknown. However, if the system
parameters depend on time adiabatically, we can seek a solution of
the Hamilton--Ehrenfest system and system in variations  in the form
of a power series in the adiabatic parameter for which $1/T$ is
taken. These solutions allow one to construct a leading term of the
asymptotics in the parameter $1/T$ of equation (\ref {pgau1}). For
solutions of this type, the Berry phase can be found in explicit
form.

\section{The spectral problem}

Consider the spectral problem (\ref{Ber2_2}) to state the Cauchy
problem (\ref{pgau1}), (\ref{Ber2_3}) and find the dynamic phase by
(\ref{phb01}).

The solution of the spectral problem can be obtained from the
non-stationary Hartree type equation (\ref{pgau1}), where the
operator $\widehat{\mathcal H}_{\varkappa}(R(t),\Psi(t))$ is
replaced by $\widehat{\mathcal H}_{\varkappa}(R,\Psi(t))$ with
$R={\rm const}$. The solutions of equation (\ref{pgau1}), which have
the form
\begin{equation}
\Psi(\vec{x},t)=\exp\Big\{-\frac{i}{\hbar}E_\nu(R)
t\Big\}\psi_\nu(\vec x,R)
\end{equation}
provide a solution of the spectral problem (\ref{Ber2_2}).  Here
$\psi_\nu(\vec{x},R)$ and
 $E_\nu(R)$ are eigenfunctions and eigenvalues of the instantaneous Hamiltonian
 $\widehat{\mathcal H}_{\varkappa}(R,\psi_\nu(R))$, respectively.

The spectral problem is related to the stationary solutions
($\dot{{\mathfrak g}}(t,{\mathfrak C})=0$) of the
Hamilton--Ehrenfest system (\ref{bbst2.6a1}) written for the
stationary nonlinear Hamiltonian $\widehat{\mathcal
H}_{\varkappa}(R,\psi)$. The solutions of (\ref{bbst2.6a1})
determine a stationary point ${\mathfrak g}(R,{\mathfrak C}_s)$ in
an extended phase space. Here by ${\mathfrak C}_s$ we have
designated a subset of constants separated from the set ${\mathfrak
C}$ by the condition of stationarity of solutions of the
Hamilton--Ehrenfest system  (see \cite{LfTS1}). From
$$
J{\mathfrak H}_z({\mathfrak g},R)=0
$$
it follows that $\vec{P}(t,{\mathfrak C}_s)=0, \vec{X}(t,{\mathfrak
C}_s)=0$.

The solutions of the Hamilton--Ehrenfest system for the second-order
moments are obtained from solutions of the system in variations
according to  (\ref{secmom}).

The linearly independent solutions of  equation (\ref{gau13q.1})
normalized by the skew-orthogonality condition (\ref{phaz1lst4.2a})
can be written
\begin{eqnarray}
&a_{\eta}(t)=e^{i\Omega_{\eta}
t}f_{\eta}(R),\qquad a_3(t)=e^{i\Omega_3 t}f_3(R);\label{gau13bb.1}\\
&f_{\eta}(R)=\left(\begin{array}{c}
\displaystyle\frac{\sqrt{m}(i\rho+\omega_a)}{{\sqrt{2\omega_a}}}(i\vec
e_{\varphi}+(-1)^{\eta}\vec
e_{\theta})\\[6pt]
\displaystyle\frac{1}{\sqrt{2m\omega_a}}(\vec
e_{\varphi}-i(-1)^{\eta}\vec e_{\theta})
\end{array}\right),\quad f_3(R)=\left(\begin{array}{c}
-\displaystyle\frac{\sqrt{m}(\rho-i\Omega_3)}{\sqrt{\Omega_3}}\vec e_n\\[6pt]
\displaystyle\frac{1}{\sqrt{m\Omega_3}}\vec e_n
\end{array}\right).\label{gau13bb}
\end{eqnarray}
Here we have used the notation
\begin{eqnarray}
&&\vec
e_n=(\cos\varphi\sin\theta,\sin\varphi\sin\theta,\cos\theta),\nonumber\\
&&\vec e_\varphi=(\sin\varphi,-\cos\varphi,0),\\
&&\vec
e_\theta=(\cos\varphi\cos\theta,\sin\varphi\cos\theta,-\sin\theta),\nonumber\\
&&
\Omega_\eta(R)=(-1)^{\eta}\displaystyle\frac{\omega_c(R)}{2}+\omega_a(R);
\qquad\Omega_3(R)
=\sqrt{\displaystyle\frac{\tilde{k}}{m}-\rho^2},\label{omeg}
\end{eqnarray}
where
\begin{eqnarray}
\eta=1,2;\qquad\omega_c(R)=\displaystyle\frac{e
H}{2mc};\qquad\omega_a(R)=\sqrt{\displaystyle\frac{e^2H^2}{4m^2c^2}+\displaystyle\frac{\tilde
k}{m}-\rho^2}.\label{omeg1}
\end{eqnarray}
The unit vector  $\vec e_n$ specifies the direction of the magnetic
field, and the set of vectors $\{\vec e_{\varphi}, \vec
e_{\theta},\vec e_{n}\}$ constitutes an orthonormal basis in
$\mathbb{R}^3$. We suppose that the friquencies
$\Omega_1,\Omega_2,\Omega_3$ do not satisfy the resonance relation
$l_1\Omega_1+l_2\Omega_2+l_3\Omega_3=0$, where $l_1,l_2,l_3$ are
integers.

The stationary solution $\Delta_2(R,{\mathfrak C}_s)$ can be shown
to exist if the solutions of the system in variations can be
presented in the form  (\ref{gau13bb.1}) (see
\cite{Sp_Bel,Bel_Kond}).

Note that, according to (\ref{gau13bb.1}), the matrices $B(t)$ and
$C(t)$ can be written
\begin{eqnarray}
&&B(t)=\tilde B(R)\Lambda(t)=G\tilde B_0(R)\Lambda(t),\qquad
C(t)=\tilde C(R)\Lambda(t)=G\tilde C_0(R)\Lambda(t),\label{ctcr}
\end{eqnarray}
where
\begin{eqnarray}
&&\tilde B_0(R)=\left(%
\begin{array}{ccc}
  -\displaystyle\frac{\sqrt{m}(\rho-i\omega_a)}{{\sqrt{2\omega_a}}}
  & -\displaystyle\frac{\sqrt{m}(\rho-i\omega_a)}{{\sqrt{2\omega_a}}} & 0 \\
-\displaystyle\frac{\sqrt{m}(i\rho+\omega_a)}{{\sqrt{2\omega_a}}}
  & \displaystyle\frac{\sqrt{m}(i\rho+\omega_a)}{{\sqrt{2\omega_a}}} & 0 \\
  0 & 0 & -\displaystyle\frac{\sqrt{m}(\rho-i\Omega_3)}{\sqrt{\Omega_3}}\\
\end{array}%
\right), \\
&& \tilde C_0(R)=\left(%
\begin{array}{ccc}
  \displaystyle\frac{1}{\sqrt{2m\omega_a}} & \displaystyle\frac{1}{\sqrt{2m\omega_a}} & 0 \\
  \displaystyle\frac{1}{\sqrt{2m\omega_a}} & \displaystyle\frac{1}{\sqrt{2m\omega_a}} & 0\\
  0 & 0 & \displaystyle\frac{1}{\sqrt{m\Omega_3}} \\
\end{array}%
\right),\\
 &&\Lambda(t)={\rm diag}\{\exp (i\Omega_1(R)t), \exp (i\Omega_2(R)t),
\exp (i\Omega_3(R)t)\}, \\
&&G=(e_\varphi, e_\theta, e_n)=\left(%
\begin{array}{ccc}
  \sin\varphi & \cos\varphi\cos\theta & \cos\varphi\sin\theta \\
  -\cos\varphi &  \sin\varphi\cos\theta & \sin\varphi\sin\theta \\
  0 & -\sin\theta & \cos\theta \\
\end{array}%
\right).
\end{eqnarray}
It is easy to verify that the matrix $G$ is orthogonal.

In view of the relations
\begin{eqnarray}
&&S(t,{\mathfrak
C}_s)=-\frac{\tilde\varkappa}2\sum_{k=1}^3\displaystyle
c\sigma_{x_kx_k}(R,{\mathfrak C}_s)t,\\
&&\det C(t)=\displaystyle\frac{-i}{m^{3/2}\Omega_3^{1/2}\omega_a}
\exp{\Big\{
{i}\Big(\Omega_1+\Omega_2+\Omega_3\Big)t\Big\}},\\
&&Q(R)=GQ_0(R)G^\intercal,\qquad
Q_0(R)={\rm diag} (im\omega_a,im\omega_a,im\Omega_3),\\
&&N_\hbar\Bigl(\det C(0)\Bigr)^{1/2}=(\pi\hbar)^{-3/4},
\end{eqnarray}
 we obtain from  (\ref{spgau13a}) the vacuum solution of the
 associated linear Schr\"odinger equation
\begin{eqnarray}
&&\Phi_0(\vec{x},t,{\mathfrak
C}_s)=\displaystyle\frac{\sqrt{i}m^{3/4}\Omega_3^{1/4}\omega_a^{1/2}}{(\pi\hbar)^{3/4}}
\exp{\Big\{\frac{i}{\hbar}\bigg(-\frac{\tilde\varkappa}2\sum_{k=1}^3\displaystyle
c\sigma_{x_kx_k}(R,{\mathfrak
C}_s)t-\displaystyle\frac{\hbar}2\sum_{k=1}^{3}\Omega_{k}t+\frac
12\langle\vec \chi, Q_0(R)\vec \chi\rangle\bigg)\Big\}}=\cr
&&=\exp{\Big\{-\frac{i}{\hbar}\bigg(\frac{\tilde\varkappa}2\sum_{k=1}^3\displaystyle
c\sigma_{x_kx_k}(R,{\mathfrak
C}_s)t+\displaystyle\frac{\hbar}2\sum_{k=1}^{3}\Omega_{k}t\bigg)\Big\}}\phi_0(\vec{x},R,{\mathfrak
C}_s). \label{gau13a}
\end{eqnarray}
Here $\vec\chi=G^{\intercal}\vec x$. In view of (\ref{phipol}) and
(\ref{ctcr}) we find
\begin{equation}
\Phi_\nu(\vec{x},t,{\mathfrak
C}_s)=\bigg(\frac{1}{\sqrt{2}}\bigg)^{|\nu|}\frac{1}{\sqrt{\nu!}}\exp\bigg\{-\displaystyle\frac
i{\hbar}\bigg(\frac{\tilde\varkappa}2\sum_{k=1}^3\displaystyle
c\sigma_{x_kx_k}(R,{\mathfrak
C}_s)t+\sum_{k=1}^{3}\hbar\Omega_{k}\Big(\nu_k+\displaystyle\frac12\Big)
t\bigg)\bigg\} \phi_0(\vec{x},R,{\mathfrak
C}_s)H_\nu\bigl(\vec{\xi}(R)\bigr) ,\label{gau19q}
\end{equation}
where  $H_\nu\bigl(\vec{\xi}(R)\bigr)$ are  Hermite polynomials,
which are set by the matrix $\widetilde{W}(R)$. Here
\begin{eqnarray}
&&\vec{\xi}(R)=\displaystyle\frac{-i}{\sqrt{\hbar}}(\widetilde{C}^*(R))^{-1}\Delta\vec{x}=
-i\sqrt {\displaystyle\frac {m}{2\hbar}}\left(%
\begin{array}{ccc}
 {\sqrt{\omega_a}} &{i\sqrt{\omega_a}} & 0 \\
  {\sqrt{\omega_a}}& {-i\sqrt{\omega_a}} &0
  \\[5 pt]
  0 & 0& \sqrt{2\Omega_3}
  \\[5 pt]
\end{array}%
\right){\vec \chi},\cr
&&\widetilde{W}(R)=-\widetilde{C}_0^{+}(R)(\widetilde{C}_0^{-1}(R))^\intercal=\left(
\begin{array}{ccc}
  0 & 1 & 0 \\
  1 & 0 & 0 \\
  0 & 0 & 1 \\
\end{array}%
\right).
\end{eqnarray}
For the considered Hermite polynomials, the following relation
results from (\ref{ctcr}):
\begin{equation}
{\rm
He}_\nu\bigl(\vec{\zeta}(R,t),t\bigr)=\exp\Big[-i\displaystyle\sum^3_{k=1}
\Omega_k\nu_kt\Big]H_\nu\bigl(\vec{\xi}(R),R\bigr).
\end{equation}

Evidently, to solve the considered spectral problem, it suffices to
know only the sub-matrix of  variances of coordinates instead of
finding  the  complete matrix of the second moments.

For the matrix of coordinate variances calculated for the states
$\Phi_\nu(\vec{x},t,{\mathfrak C}_s)$, the following formula is
valid
 (see, e.g.,  \cite{Bagre}):
\begin{equation}
\sigma_{xx}=\displaystyle\frac{\hbar}{4}[\tilde C(R)D_\nu^{-1}\tilde
C^+(R)+\tilde C^*(R)D_\nu^{-1}\tilde C^\intercal(R)],\qquad
D_\nu^{-1}=\Big\|(2\nu_j+1)\delta_{kj} \Big\|_{3\times
3}.\label{sigxx}
\end{equation}
Using (\ref{sigxx}), we find
\begin{eqnarray}
&&\sum_{k=1}^3\sigma_{x_kx_k}(R,{\mathfrak C}_s)=\bigg({\nu}_1+
\frac{1}{2}\bigg)\displaystyle\frac{{\hbar}}{m\omega_a}+\bigg({\nu}_2+
\frac{1}{2}\bigg)\displaystyle\frac{{\hbar}}{m\omega_a}+\bigg({\nu}_3+
\frac{1}{2}\bigg)\displaystyle\frac{{\hbar}}{m\Omega_3}.
\end{eqnarray}
Then the solution of equation \eqref{pgau1} is
\begin{eqnarray}
&&\Psi(\vec{x},t)=\Phi_\nu(\vec{x},t,\overline{\mathfrak
C}_\nu)=\bigg(\frac{1}{\sqrt{2}}\bigg)^{|\nu|}\frac{1}{\sqrt{\nu!}}\exp\bigg\{-i\sum_{k=1}^{3}
(\Omega_{k}+\tilde\Omega_{k})\bigg(\nu_k+\displaystyle\frac12\bigg)
t\bigg\} H_\nu\bigl(\vec{\xi}(R)\bigr)
\phi_0(\vec{x},R,\overline{\mathfrak C}_\nu),\label{gau19qa}\cr &&
\tilde\Omega_{1}(R)=\tilde\Omega_{2}(R)=\displaystyle\frac{\tilde\varkappa
c}{2m\omega_a(R)},\quad\tilde\Omega_{3}(R)=\displaystyle\frac{\tilde\varkappa
c}{2m\Omega_3(R)}.\label{gau13abc}
\end{eqnarray}
Therefore, the eigenfunctions of the Hartree operator (\ref{phbb07})
are
\begin{eqnarray}
&&\psi_{\nu}(\vec{x},R)=\Big(\frac{1}{\sqrt{2}}\Big)^{|\nu|}\frac{1}{\sqrt{{\nu}!}}
\displaystyle\frac{\sqrt{i}m^{3/4}\Omega_3^{1/4}\omega_a^{1/2}}{(\pi\hbar)^{3/4}}
\exp{\Big\{\frac{i}{\hbar}\Big(\frac 12\langle\vec \chi, Q_0(R)\vec
\chi\rangle\Big)\Big\}}{H}_{\nu}\bigl(\vec{\xi}(R)\bigr),
\label{gau13ab}
\end{eqnarray}
and the corresponding eigenvalues are given by the expression
\begin{eqnarray}
&&E_{\nu}(R)=\hbar\sum_{k=1}^{3}
(\Omega_{k}(R)+\tilde\Omega_{k}(R))\Big(\nu_k+\displaystyle\frac12\Big).
\end{eqnarray}

\section{Adiabatic approximation and the Berry phase}

Assume that the evolution of a quantum system goes adiabatically.
This implies that the Hamiltonian parameters slowly change with time
[see (\ref{phb07})] and, along with the ``fast'' time $t$ entering
the time derivative operator, a ``slow'' time $s$ can be introduced
which the Hamiltonian parameters ($R(t)$=$R(s)$) depend on. Let the
``fast'' and the ``slow'' time are related as
\begin{equation}
s=t/T,\label{adiab01}
\end{equation}
where $T$ is the evolution period of the system.

As mentioned above,  to find a solution of  equation (\ref {pgau1})
in the adiabatic approximation, it is necessary to solve the
Hamilton--Ehrenfest system  and  system in variations accurate to
the second order in $1/T $.

The Hamilton--Ehrenfest system can be written as
\begin{equation}
 \left\{
\begin{array}{l}\displaystyle\frac1T z_{\Psi}' = J{\mathfrak H}_z({\mathfrak g}_\Psi,R(s)),\\[8pt]
 \displaystyle\frac1T\Delta_{2\Psi}'=J{\mathfrak H}_{zz}(R(s))\Delta_{2\Psi}-\Delta_{2\Psi}{\mathfrak H}_{zz}(R(s))J,
 \quad\Delta_2^\intercal=\Delta_2,
\end{array}\right.\label{bbst2.6ad}
\end{equation}
where $a'=da/ds$. We seek a solution of this system in the form
\begin{eqnarray}
\vec{Z}(t)=\vec{Z}^{(0)}(s)+\frac{1}{T}\vec{Z}^{(1)}(s)+O\bigg(\displaystyle\frac1{T^2}\bigg),\qquad
\Delta_2(t)=\Delta_2^{(0)}(s)+\frac{1}{T}\Delta_2^{(1)}(s)+O\bigg(\displaystyle\frac1{T^2}\bigg),
\label{adiab04}
\end{eqnarray}
and obtain
\begin{eqnarray}
&&\vec{X}^{(0)}(s)=\vec{X}^{(1)}(s)=0,\qquad \vec P^{(0)}(s)=\vec
P^{(1)}(s)=0. \label{adiab05}
\end{eqnarray}
As in the spectral problem, we obtain the solution of the
Hamilton--Ehrenfest system for the second order moments using the
solutions of the system in variations.

Making the change of  variables by (\ref{adiab01}) in the system in
variations (\ref{gau13q.1}), we obtain
\begin{equation}
\frac{1}{T}a'(t)=J{\mathfrak H}_{zz}(s)a(t). \label{adgau13}
\end{equation}
Let us seek a semiclassical asymptotic solution of the system
(\ref{adgau13}) in the form
\begin{equation}
a_k(t)=e^{i(T\Phi_k(s)+\phi_k(s))}f_k(t)+O\bigg(\displaystyle\frac1{T^2}\bigg),\label{adiab08}
\end{equation}
\begin{equation}
f_k(t)=f_k^{(0)}(s)+\frac{1}{T}f_k^{(1)}(s),\quad k=\overline{1,3}.
\end{equation}
Substituting (\ref{adiab08}) in (\ref{adgau13}) and equating the
coefficients of equal powers of  $1/T$, we obtain
\begin{eqnarray}
&&(J{\mathfrak H}_{zz}(s)-i\Phi_k'(s))f_k^{(0)}(s)=0,\cr
&&(J{\mathfrak
H}_{zz}(s)-i\Phi_k'(s))f_k^{(1)}(s)=f_k^{(0)'}(s)+i\phi_k'(s)f_k^{(0)}(s).
\end{eqnarray}
Then
\begin{eqnarray}
&&\Phi_k'(s)=\Omega_k(s)=\Omega_k(R(s)),\quad k=\overline{1,3} \\[8 pt]
&&f_{\eta}^{(0)}(s)=f_{\eta}(R(s)),\quad
f_3^{(0)}(s)=f_3(R(s)),\quad \eta=1,2, \label{gau13bbq}
\end{eqnarray}
where the vectors $f_{\eta}(R(s))$, $f_3(R(s))$ and the functions
 $\Omega_k(R(s))$ are determined by (\ref{gau13bb}) and (\ref{omeg}), respectively.

Let us decompose the vectors  $f_k^{(1)}(s)$ in the basis vectors
$f_k^{(0)}(s)$ and $f_k^{(0)*}(s)$:
\begin{equation}
f_k^{(1)}(s)=\sum_{m=1}^{3}\alpha_{km}(s) f_m^{(0)}(s)+\beta_{km}(s)
f_m^{(0)*}(s).
\end{equation}
Then we obtain
\begin{eqnarray}
&\phi_k'(s)=\displaystyle\frac{1}{2}\{f_k^{(0)'}(s),f_k^{(0)*}(s)\},\\
&\alpha_{km}(s)=\displaystyle\frac{\{f_k^{(0)'}(s),f_m^{(0)*}(s)\}}{2(\Omega_k(s)-\Omega_m(s))},\quad
\beta_{km}(s)=\displaystyle\frac{\{f_m^{(0)}(s),f_k^{(0)'}(s)\}}{2(\Omega_m(s)+\Omega_k(s))}.
\end{eqnarray}
Note that
\begin{eqnarray}
&&\alpha_{km}(s)=-\alpha_{mk}^{*}(s),\qquad
\beta_{km}(s)=\beta_{mk}(s)\,
\end{eqnarray}
or, in matrix form,
\begin{eqnarray}
&&\bf A(s)=-\bf A^{+}(s),\qquad \bf B(s)=\bf B^\intercal(s),
\end{eqnarray}
where ${\bf A(s)}=\|\alpha_{km}(s)\|$ and ${\bf
B(s)}=\|\beta_{km}(s)\|$. By analogy with (\ref{ctcr}), we write
\begin{eqnarray}
&&B(t)=\big(\tilde B^{(0)}(s)+\frac1{T}\tilde
B^{(1)}(s)\big)\Lambda^{(0)}(s,T)+O\bigg(\displaystyle\frac1{T^2}\bigg),\cr
&&C(t)=\big(\tilde C^{(0)}(s)+\frac1{T}\tilde
C^{(1)}(s)\big)\Lambda^{(0)}(s,T)+O\bigg(\displaystyle\frac1{T^2}\bigg),\label{ctcr.1}\\
&&\Lambda^{(0)}(s,T)={\rm diag}\{T\Phi_1(s)+\phi_1(s),
T\Phi_2(s)+\phi_2(s), T\Phi_3(s)+\phi_3(s)\}.\nonumber
\end{eqnarray}
Here
\begin{eqnarray}
&&\tilde B^{(0)}(s))=\tilde B(R(s)),\qquad \tilde C^{(0)}(s)=\tilde
C(R(s)),\cr &&\tilde B^{(1)}(s)=\tilde B^{(0)}(s){\bf
A}^\intercal(s)+\tilde B^{(0)*}(s){\bf B}^\intercal(s),\label{bc1}\\
&&\tilde C^{(1)}(s)=\tilde C^{(0)}(s){\bf A}^\intercal(s)+\tilde
C^{(0)*}(s){\bf B}^\intercal(s).\nonumber
\end{eqnarray}
The matrices  $\tilde B^{(0)}(s)$ and $\tilde C^{(0)}(s)$ are
determined by the vectors (\ref{gau13bbq}). Hereinafter we omit the
argument $s$ or $t$ if this  does not lead to confusion.

Note that
\begin{eqnarray}
&\phi_1'=\displaystyle\frac{-(m\rho)'}{2m\omega_a}+\displaystyle\frac12\big(\langle
e_\varphi',e_\theta\rangle - \langle
e_\theta',e_\varphi\rangle\big),\qquad
\phi_2'=\displaystyle\frac{-(m\rho)'}{2m\omega_a}+\displaystyle\frac12\big(\langle
e_\theta',e_\varphi\rangle-\langle
e_\varphi',e_\theta\rangle\big),\qquad\phi_3'=\displaystyle\frac{-(m\rho)'}{2m\Omega_3};\cr
&Q(t)={\rm diag}
(im\omega_a(s),im\omega_a(s),im\Omega_3(s))+O\bigg(\displaystyle\frac1{T}\bigg).
\end{eqnarray}
For the matrices  $\bf A$ and $\bf B$ we obtain
\begin{eqnarray}
\bf A=\left(%
\begin{array}{ccc}
  \alpha_{11} & 0 &
  \displaystyle\frac{\gamma}{2(\Omega_1-\Omega_3)}\langle e_n',(-ie_\varphi+e_\theta)\rangle\\
 0 & \alpha_{22} & \displaystyle\frac{\gamma}{2(\Omega_2-\Omega_3)}\langle e_n',(-ie_\varphi-e_\theta)\rangle \\
  \displaystyle\frac{\gamma}{2(\Omega_3-\Omega_1)}\langle e_n',(ie_\varphi+e_\theta)\rangle &
  \displaystyle\frac{\gamma}{2(\Omega_3-\Omega_2)}\langle e_n',(ie_\varphi-e_\theta)\rangle & \alpha_{33} \\
\end{array}%
\right);\nonumber\\[10 pt]
\bf B=\left(%
\begin{array}{ccc}
  0 &
   \displaystyle\frac{(m\rho-im\omega_a)'}{4m\omega_a^2}
   & \displaystyle\frac{\tilde\gamma}{2(\Omega_3+\Omega_1)}\langle e_n',(-ie_\varphi+e_\theta)\rangle \\
 \displaystyle\frac{(m\rho-im\omega_a)'}{4m\omega_a^2}  &
   0 & \displaystyle\frac{\tilde\gamma}{2(\Omega_3+\Omega_2)}\langle e_n',(-ie_\varphi-e_\theta)\rangle  \\
  \displaystyle\frac{\tilde\gamma}{2(\Omega_3+\Omega_1)}\langle e_n',(-ie_\varphi+e_\theta)\rangle&
  \displaystyle\frac{\tilde\gamma}{2(\Omega_3+\Omega_2)}\langle e_n',(-ie_\varphi-e_\theta)\rangle  & \displaystyle\frac{(m\rho-im\Omega_3)'}{4m\Omega_3^2} \\
\end{array}%
\right).\nonumber
\end{eqnarray}
Here
\begin{eqnarray}
\gamma=\bigg[\sqrt{\displaystyle\frac{\Omega_3}{2\omega_a}}+\sqrt{\displaystyle\frac{\omega_a}{2\Omega_3}}\bigg],\qquad
\tilde\gamma=\bigg[\sqrt{\displaystyle\frac{\Omega_3}{2\omega_a}}-\sqrt{\displaystyle\frac{\omega_a}{2\Omega_3}}\bigg];
\end{eqnarray}
the matrix elements  $\alpha_{11}$, $\alpha_{22}$, and $\alpha_{33}$
are imaginary, and they are determined from the next-order
approximation. We do not give them in explicit form as these
functions do not contribute to the leading term of the asymptotic
expansion.

To construct solutions of the nonlinear equation (\ref{pgau1}), we
need the matrix  of coordinate variances
\begin{eqnarray}
\sigma_{xx}(t)=\displaystyle\frac{\hbar}{4}\big[C(t)D_{\nu}^{-1}C^+(t)+C^{*}(t)D_{\nu}^{-1}C^{T}(t)\big],
\end{eqnarray}
and it suffices to know this matrix accurate to  $O(1/T^2)$:
\begin{eqnarray}
\sigma_{xx}(t)=\sigma_{xx}^{(0)}(s)+\displaystyle\frac{1}{T}\sigma_{xx}^{(1)}(s)+
O\bigg(\displaystyle\frac1{T^2}\bigg).
\end{eqnarray}
Here
\begin{eqnarray}
&&\sigma_{xx}^{(0)}= \displaystyle\frac{\hbar}{4}\big[\tilde
C^{(0)}D_{\nu}^{-1}\tilde C^{(0)+}+\tilde
C^{(0)*}D_{\nu}^{-1}\tilde C^{(0)\intercal}\big],\\
&&\sigma_{xx}^{(1)}=\displaystyle\frac{\hbar}{4}\big[\tilde
C^{(0)}D_{\nu}^{-1}\tilde{C}^{(1)+}+\tilde
C^{(1)}D_{\nu}^{-1}\tilde{C}^{(0)+}+\tilde{C}^{(0)*}D_{\nu}^{-1}\tilde{C}^{(1)T}+
\tilde{C}^{(1)*}D_{\nu}^{-1}\tilde{C}^{(0)T}\big]=\cr&&
=\displaystyle\frac{\hbar}{4}\big[\tilde C^{(0)}\{{\bf
A}^{\intercal}D_{\nu}^{-1}-D_{\nu}^{-1}{\bf
A}^{\intercal}\}\tilde{C}^{(0)+}+\tilde C^{(0)}\{D_{\nu}^{-1}{\bf
B}^{*}+{\bf B}^{*}D_{\nu}^{-1}\}\tilde{C}^{(0)\intercal}+\cr
&&+\tilde C^{(0)*}\{D_{\nu}^{-1}{\bf A}-{\bf
A}D_{\nu}^{-1}\}\tilde{C}^{(0)\intercal}+\tilde
C^{(0)*}\{D_{\nu}^{-1}{\bf B}+{\bf B}
D_{\nu}^{-1}\}\tilde{C}^{(0)+}\big].
\end{eqnarray}
The leading term of the asymptotics depends only on the sum of the
diagonal elements of this matrix
\begin{eqnarray}
&&\sum_{k=1}^3\sigma_{x_kx_k}(t)=\bigg({\nu}_1+
\frac{1}{2}\bigg)\bigg[\displaystyle\frac{{\hbar}}{m\omega_a}+
\displaystyle\frac1T\displaystyle\frac{\hbar(m\rho)'}{2m^2\omega_a^3}\bigg]+\cr&&+\bigg({\nu}_2+
\frac{1}{2}\bigg)\bigg[\displaystyle\frac{{\hbar}}{m\omega_a}+
\displaystyle\frac1T\displaystyle\frac{\hbar(m\rho)'}{2m^2\omega_a^3}\bigg]+\bigg({\nu}_3+
\frac{1}{2}\bigg)\bigg[\displaystyle\frac{{\hbar}}{m\Omega_3}+
\displaystyle\frac1T\displaystyle\frac{\hbar(m\rho)'}{2m^2\Omega_3^3}\bigg]+
O\bigg(\displaystyle\frac1{T^2}\bigg).\nonumber
\end{eqnarray}
In view of this, a solution of the Cauchy problem
 (\ref{pgau1}), (\ref{Ber2_3}) can be represented in the form
\begin{equation} \Psi(\vec{x},t)=\Psi_{\nu}^{(0)}(\vec{x},t)+O\bigg(\frac1T\bigg),
\end{equation}
where
\begin{equation}
\Psi_{\nu}^{(0)}(\vec{x},t)=\exp\bigg\{-\frac{i}{\hbar}T\int\limits_{0}^{s}E_\nu(\tau)d\tau+
i\gamma_{\nu}(s)\bigg\}\psi_{\nu}(\vec{x},R(s)).\label{adiab10}
\end{equation}
The function  (\ref{adiab10}) is a solution of equation
(\ref{pgau1}) in the adiabatic approximation. The quantities
\begin{equation}
E_{\nu}(s)=\hbar\sum_{j=1}^{3}
(\Omega_{j}(s)+\tilde\Omega_{j}(s))\bigg(\nu_j+\displaystyle\frac12\bigg)
\end{equation}
are eigenfunctions of the instantaneous Hartree type Hamiltonian
$\hat{\mathcal H}_{\varkappa}(R(s),\psi_\nu(R(s)))$, and
$\gamma_{\nu}(s)$ is
\begin{eqnarray}
&&\gamma_{\nu}(s)=-\int\limits_{0}^{s}\sum_{j=1}^{3}
\bigg[\phi'_j(\tau)\bigg(\nu_j+\displaystyle\frac12\bigg)+\frac{\tilde\varkappa
c(\tau)}{2\hbar}\sum_{j=1}^{3}\sigma_{x_jx_j}^{(1)}(\tau)\bigg]d\tau=\cr
&&=\sum_{j=1}^{2}\bigg(\nu_j+\displaystyle\frac12\bigg)\int\limits_{0}^{s}
\bigg[1-\displaystyle\frac{\tilde\varkappa
c(\tau)}{2m(\tau)\omega_a^2(\tau)}\bigg]
\displaystyle\frac{(m(\tau)\rho(\tau))'}{2m(\tau)\omega_a(\tau)}d\tau+
\bigg(\nu_3+\displaystyle\frac12\bigg)\int\limits_{0}^{s}\bigg[1-\displaystyle\frac{\tilde\varkappa
c(\tau)}{2m(\tau)\Omega_3^2(\tau)}\bigg]
\displaystyle\frac{(m(\tau)\rho(\tau))'}{2m(\tau)\Omega_3(\tau)}d\tau+\cr
&&+\displaystyle\frac12\Big(\nu_2-\nu_1\Big)\int\limits_{0}^{s}{\langle
e_\varphi'(\tau),e_\theta(\tau)\rangle\ - \langle
e_\theta'(\tau),e_\varphi(\tau)\rangle}d\tau.
\end{eqnarray}
The functions $\Omega_{j}(s)$, $\omega_a(s)$, and
$\tilde\Omega_{j}(s)$ are defined by (\ref{omeg}), (\ref{omeg1}),
and (\ref{gau13abc}), respectively. The evolution of the function
(\ref{adiab10}) in the period yields
\begin{equation}
\Psi_{\nu}^{(0)}(\vec{x},T)=\exp\bigg\{-\frac{i}{\hbar}T\int\limits_{0}^{1}E_\nu(s)ds+
i\gamma_{\nu}(T)\bigg\}\Psi_{\nu}^{(0)}(\vec{x},0).
\end{equation}
We now use (\ref{Ber2_5}) and   (\ref{phb01}) to define the dynamic
phase
\begin{equation}
\delta_{\nu}(T)=T\int\limits_{0}^{1}\hbar\sum_{j=1}^{3}
\bigg(\nu_j+\displaystyle\frac12\bigg)\big[\Omega_{j}(s)+\tilde\Omega_{j}(s)\big]ds\label{adiab12}
\end{equation}
and the Berry phase
\begin{eqnarray}
&&\gamma_{\nu}(T)=\sum_{j=1}^{2}\bigg(\nu_j+\displaystyle\frac12\bigg)\oint\limits_{C}
\bigg[1-\displaystyle\frac{\tilde\varkappa c}{2m\omega_a^2}\bigg]
\displaystyle\frac{1}{2\omega_a}\bigg(\displaystyle\frac{\rho}{m}dm+d\rho\bigg)+
\bigg(\nu_3+\displaystyle\frac12\bigg)\oint\limits_{C}\bigg[1-\displaystyle\frac{\tilde\varkappa
c}{2m\Omega_3^2}\bigg]\displaystyle\frac{1}{2\Omega_3}\bigg(\displaystyle\frac{\rho}{m}dm+d\rho\bigg)+\cr
&&+\Big(\nu_2-\nu_1\Big)\oint\limits_{C}\displaystyle\frac{H_3}{H(H_1^2+H_2^2)}[H_1dH_2-H_2dH_1].
\label{adiab13}
\end{eqnarray}
Here, $H_1$, $H_2$, and $H_3$ are components of the magnetic field,
$H=\sqrt{H_1^2+H_2^2+H_3^2}$; $C$ is a contour in the parameter
space.

The Berry phase (\ref{adiab13}) of the nonlinear equation
(\ref{pgau1}) differs from that of  the linear Schr\"odinger
equation ($\tilde\varkappa=0$) by the presence of the parameter
$\tilde k$ instead of $k$ and by an additional summand proportional
to $\tilde\varkappa$.

Let $a=c$ in (\ref{V}). Then for $(\Omega_{\rm nl}/\Omega_0)^2<<1$
we obtain
\begin{eqnarray}
&&\gamma_{\nu}(T)=\sum_{j=1}^{2}\bigg(\nu_j+\displaystyle\frac12\bigg)\oint\limits_{C}
\bigg[1-\bigg(\displaystyle\frac{\Omega_{\rm
nl}}{\omega_0}\bigg)^2\bigg]
\displaystyle\frac{1}{2\omega_0}\bigg(\displaystyle\frac{\rho}{m}dm+d\rho\bigg)+\label{adiab13.1}\\
&&+ \bigg(\nu_3+\displaystyle\frac12\bigg)
\oint\limits_{C}\bigg[1-\bigg(\displaystyle\frac{\Omega_{\rm
nl}}{\Omega_0}\bigg)^2\bigg]
\displaystyle\frac{1}{2\Omega_0}\bigg(\displaystyle\frac{\rho}{m}dm+d\rho\bigg)
+\Big(\nu_2-\nu_1\Big)\oint\limits_{C}\displaystyle\frac{H_3}{H(H_1^2+H_2^2)}[H_1dH_2-H_2dH_1],\cr
&& \Omega_{\rm nl}=\sqrt{\displaystyle\frac{\tilde\varkappa
a}{m}},\quad
\Omega_0=\Omega_3\big|_{\tilde\varkappa=0}=\sqrt{\displaystyle\frac{k}{m}-\rho^2},
\quad\omega_0=\omega_a\big|_{\tilde\varkappa=0}=
\sqrt{\displaystyle\frac{e^2H^2}{4m^2c^2}+\displaystyle\frac{k}{m}-\rho^2}.
\end{eqnarray}

The classical analog of the Berry phase is associated with the
Hannay angles  (see, e.g.,  \cite{Hannay1}). Geometrically, the
Hannay angles similar to the Berry phases. The Berry phase
$\gamma_\nu$ of a quantum system and the Hannay angles $\Theta_i$ of
the corresponding classical system \cite{Berry85} are related by
 \begin{eqnarray}
\Theta_i=-\hbar\frac{\partial\gamma_\nu}{\partial
I_i}=-\frac{\partial\gamma_\nu}{\partial \nu_i},\quad
i=\overline{1,3}. \label{adgauh}
\end{eqnarray}
Here  $I_i$ are  quantized action variables and  $\nu_i$ are quantum
numbers. The differentiation in (\ref{adgauh}) implies that
 $\nu_i$ is a continuous parameter. In the nonlinear case, it is  naturally associate
 the the Berry phase (\ref{adiab13}) with an analog of the Hannay angle
 by formula (\ref{adgauh}):
\begin{eqnarray}
&&\Theta^{\varkappa}_1=-\oint
\bigg[1-\displaystyle\frac{\tilde\varkappa c}{2m\omega_a^2}\bigg]
\displaystyle\frac{1}{2\omega_a}\bigg(\displaystyle\frac{\rho}{m}dm+d\rho\bigg)
+\oint\limits_{C}\displaystyle\frac{H_3}{H(H_1^2+H_2^2)}[H_1dH_2-H_2dH_1],\\
&&\Theta^{\varkappa}_2=-\oint
\bigg[1-\displaystyle\frac{\tilde\varkappa c}{2m\omega_a^2}\bigg]
\displaystyle\frac{1}{2\omega_a}\bigg(\displaystyle\frac{\rho}{m}dm+d\rho\bigg)
-\oint\limits_{C}\displaystyle\frac{H_3}{H(H_1^2+H_2^2)}[H_1dH_2-H_2dH_1],\cr
&&\Theta^{\varkappa}_3=\oint\bigg[1-\displaystyle\frac{\tilde\varkappa
c}{2m\Omega_3^2}\bigg]\displaystyle\frac{1}{2\Omega_3}\bigg(\displaystyle\frac{\rho}{m}dm+d\rho\bigg).\nonumber
\end{eqnarray}
When $\rho=0$, the ``Hannay angle'' \, $\Theta_1^{\varkappa}$ is
equal to the solid angle $\Omega(\tilde C)$ under which the curve
$\tilde C$ can be observed from the origin point in the
three-dimensional parameter space $(H_1, H_2, H_3)$.

We have defined the Hannay angle in terms of quantum mechanics,
since the nonlinear problem requires a special study of the
``classical equations''\, corresponding to the nonlinear
``quantum''\, Hartree type equation. In our consideration, the part
of these classical equations is played by the Hamilton--Ehrenfest
system (\ref{bbst2.6a1}), which has no Hamiltonian form with respect
to the standard Poisson bracket.

\section{Conclusion}

A countable set of solutions has been constructed for the Hartree
type equation  (\ref{phbb07}) in the adiabatic approximation and the
corresponding Berry phases have been found in explicit form.

In the linear case  ($\tilde\varkappa=0$), the Berry phase is
determined completely by the solution of the Hamilton system and the
corresponding complex germ \cite{ty94}. In the nonlinear case, the
Hamilton--Ehrenfest system involves the Hamiltonian system and the
complex germ. For the quadratic Hartree type operator
(\ref{phbb07}), the Hamilton--Ehrenfest system
 (\ref{bbst2.6a1}) breaks into the Hamiltonian system
  with a self-action (the first equation in (\ref{bbst2.6a1})) and the system in variations
(\ref{matvar}).

 It should be noted that, as in the linear case,
 the leading  term of the asymptotics (\ref{adiab10})
 in the parameter $1/T $ remains  an eigenfunction
 of the nonlinear Hamiltonian (\ref{phbb07}) at every point in time
 and  only gains  a phase factor.  However, the  Berry phase
 for the nonlinear Hartree type equation (\ref{pgau1}) cannot
 be calculated by formula  (\ref{phb04}).
 This is because  the linear principle
 of superposition was essentially used in deriving formula (\ref{phb04}).
 The Berry phase is obtained by selecting
the dynamic part (by formulas (\ref {Ber2_5}), (\ref {phb01})) from
the total phase incursion in the period $T $.

There exists a limit as $\tilde\varkappa\to 0$ for every expression
obtained,
 and the limits agree with results of linear quantum theory
(see, e.g.,  \cite{ty94,Moore,Biswas}).

The geometric Aharonov--Anandan phase \cite{shapovalov:ANANDAN} is a
generalization of  the Berry phase for the case of arbitrary cyclic
states $(\Psi(T)=\exp[\Phi(T)]\Psi(0))$.

The geometric Aharonov--Anandan  phase for the linear Schr\"odinger
equation coincides with the Berry phase in the limiting case of
adiabatic evolution $(T\to \infty)$ \cite{Biswas}.

The generalization of the  Aharonov--Anandan phase for cyclic
solutions of nonlinear equations  with unitary nonlinearity, to
which class  equation (\ref {pgau1}) also belongs, is given in
\cite{Garrison}.

In the case  of adiabatic evolution, the limit of the
Aharonov--Anandan phase apparently, depends on the type of the
nonlinear equation and on the class of functions in which its
solutions are sought. The relationship between the Aharonov-Anandan
phase as $T\to \infty $ and the Berry phase  for solutions of the
nonlinear Hartree type equation  (\ref {pgau1}) in the  class of
functions (\ref {gau4}) can be a subject of  further research.

From the view  point of  practical applications in quantum
computations, the non-Abelian Berry phase is important
\cite{Rasetti, Pachos} which corresponds  to a degenerated spectrum
of the Hamiltonian. The method developed in the work can be easily
generalized  for non-Abelian phases. Note that, in contrast to the
linear case, the occurrence of a non-Abelian phase is not related to
the degeneration of the spectrum of the nonlinear operator,  but it
is related to degeneration of the spectrum of the corresponding
associated linear operator. A complete investigation of this problem
is beyond the scope of this work and requires special consideration.

\subsection*{Acknowledgements}
The work was supported in part by  President of the Russian
Federation, Grant  No SS-5103.2006.2, DFG, Litvinets~F. was
supported in part by the scholarship of the nonprofit Dynasty
Foundation.

\end{document}